\documentstyle[prd,aps,floats,twocolumn,tighten,epsfig]{revtex}

\begin{document}

\twocolumn[\hsize\textwidth\columnwidth\hsize\csname @twocolumnfalse\endcsname
\title{Kinetic decoupling of neutralino dark matter}
\author{Xuelei Chen\thanks{}}
\address{Department of Physics, The Ohio State University,
174 West 18th Avenue, Columbus, Ohio 43210, USA.}
\author{Marc Kamionkowski\thanks{}}
\address{California Institute of Technology,
Mail Code 130-33, Pasadena, CA
91225, USA.}
\author{Xinmin Zhang\thanks{}}
\address{Institute of High Energy Physics, Chinese Academy of Sciences,
P.O.Box 918-4, Beijing 100039, P. R. China.
}

\date{\today}

\maketitle

\begin{abstract}
After neutralinos cease annihilating in the early Universe, they 
may still scatter elastically from other particles in the
primordial plasma.  At some point in time, however, they will
eventually stop scattering.  We calculate the cross sections for
neutralino elastic scattering from standard-model particles to
determine the time at which this kinetic decoupling occurs.
We show that kinetic decoupling occurs above a temperature
$T\sim$MeV.  Thereafter, neutralinos act as collisionless cold
dark matter.
\end{abstract}

]

\section{Introduction}

Neutralinos provide perhaps the most promising candidate for the 
mysterious dark matter in galactic halos (see, e.g.,
Refs. \cite{report,bergstrom} for reviews).  Such particles would have
existed in thermal equilibrium in the early Universe when the
temperature exceeded the mass of the particle.  However, as the
temperature drops below the mass, the rate of the
production and destruction reactions that maintain the
equilibrium density become smaller than the expansion rate.  At
this point, annihilations cease, and a relic population of neutralinos
remains.  The calculation of this relic density is
straightforward, and to a first approximation, the relic
density is fixed by the cross section for neutralino
annihilation to lighter standard-model particles.  The cross
section required for the critical density is $\sim10^{-37}$
cm$^2$, coincidentally within a few orders of magnitude of the
neutralino annihilation cross section in the minimal
supersymmetric extension of the standard model.

After freezeout (i.e., after they fall out of chemical
equilibrium) neutralinos may still continue scattering 
elastically from other particles in the primordial plasma, and
thus remain in chemical equilibrium.  In an interesting recent paper
\cite{Boehm}, B\oe hm et al. pointed out that if these particles
remain in kinetic equilibrium until a temperature $T\sim0.1$ keV, at 
which point the mass enclosed in a Hubble volume is
comparable to that of a dwarf galaxy, then collisional damping due to
elastic scattering of neutrinos and/or photons from neutralinos
could erase structure on such scales and thus help resolve the
discrepancy between the predicted and observed substructure in
Galactic halos\cite{moore,klypin}.  
They also argued briefly that supersymmetric
dark-matter candidates may have the required neutralino-photon
and/or neutralino-neutrino cross sections.

When heavy neutrinos were first considered as cold dark matter 
candidates, Gunn et al. \cite{gunn} showed that they would fall
out of kinetic equilibrium shortly after their annihilation 
freezes out. Later, Schmid et al. \cite{ssw} estimated the 
decoupling temperature of neutralino-neutrino interaction.
In this paper we consider the process of kinetic decoupling of
neutralinos in more detail.  We calculate the cross sections for 
elastic scattering of neutrinos and photons from neutralinos,
and for completeness, we consider the elastic scattering of
neutralinos from other particles as well.  We show that the
estimates for neutrino-neutralino and photon-neutralino
scattering used by B\oe hm et al. neglect important kinematic
effects, and thus hugely overestimate these cross sections.  We
show that neutralinos are more likely to undergo kinetic
freezeout much earlier, at temperatures between an MeV and a GeV,
rather than a keV. Thus, although B\oe hm et al.'s mechanism for
collisional damping may be applicable to more general
dark-matter candidates, it is unlikely to be relevant to the
most promising supersymmetric dark-matter candidates.

In the next Section we briefly review the cross sections
required for the B\oe hm et al. collisional damping.  In Section
III, we calculate the cross section for neutralino-photon
scattering.  In Section IV, we move on to the
neutralino-neutrino cross section.  We consider elastic
scattering of neutralinos from other species in Section V, and
we conclude in Section VI.

\section{REQUIRED CROSS SECTIONS}

In the standard scenario,
neutrinos undergo chemical decoupling at a temperature of an MeV 
and their kinetic decoupling follows shortly thereafter.
Now suppose that the neutrino has a nonzero elastic-scattering
cross section with the dark-matter particle $\chi$.  This
elastic scattering will keep the dark-matter particles in kinetic
equilibrium until the rate at which neutrinos scatter from a
given dark-matter particle becomes comparable to the expansion
rate.  The scattering rate is $\Gamma=n_\nu \sigma_{\nu\chi}c$,
where $n_\nu$ is the number density of neutrinos and
$\sigma_{\nu\chi}$ is the neutrino--dark-matter
elastic-scattering cross section.  The Hubble expansion rate is
$H=1.66 g_*^{1/2} T^2/m_{\rm Pl}$, where $g_*=3.36$ is the effective
number of relativistic degrees of freedom at a temperature
$T\sim$keV, and $m_{\rm Pl}\simeq10^{19}$ GeV is the Planck
mass.  The temperature at which the Hubble volume encloses a
baryonic mass $M_b$ is 
\begin{equation}
     T=0.15\, {\rm keV}\, (M_b/10^{9}\, M_\odot)^{-1/3}
     (\Omega_b h^2/0.02)^{1/3},
\end{equation}
where $\Omega_b$ is the baryon density, $h$ is the present Hubble
parameter in units of 100 km~sec$^{-1}$~Mpc$^{-1}$.
Setting $\Gamma=H$, at $T\sim0.1$ keV, we find that the cross
section required to erase structure on $\lesssim10^9\, M_\odot$
scales is $\sigma_{\nu\chi}\sim 10^{-40}$ cm$^2$,
which recovers the estimate
of B\oe hm et  al.  Similar arguments for scattering from 
photons lead B\oe hm et al. to estimate a dark-matter--photon
cross section of $10^{-38}$ cm$^2$ for collisional damping via
photon scattering.

In fact, we believe that B\oe hm et al. may have underestimated
the required cross section by several orders of magnitude.
The cold-dark-matter particles are non-relativistic,
so even if neutrino and photon scattering keep them in kinetic 
equilibrium, they could hardly move far enough
to smooth out fluctuations in the dark-matter density.
For collisional damping to work, the motion of photons and
neutrinos must be affected. Indeed, this is how collisional damping
works in the case of baryons (Silk damping) in which density
perturbations are smoothed on scales small compared with the
photon diffusion length\cite{kt}.
If it is in fact the neutrinos (or photons) that are responsible 
for collisional damping with neutralino-neutrino  
scattering, then the
appropriate reaction rate to equate to the Hubble length is the
rate at which a given neutrino scatters from a neutralino,
rather than {\it vice versa}.  In this case, the required
neutrino--dark-matter cross section must be larger by the ratio
of  the neutrino number density to the dark-matter number
density, and this factor is $5\times10^7\, (m_\chi/{\rm GeV})$,
where $m_\chi$ is the dark-matter-particle mass.  

\section{Neutralino-photon scattering}

\begin{figure}
\begin{center}
\epsfig{file=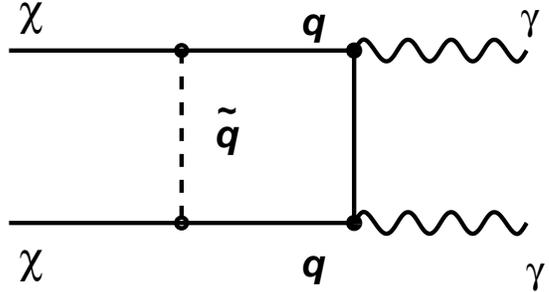,width=0.4\textwidth}
\vspace{4pt}
\caption{
One of the Feynman diagrams for neutralino annihilation to two
photons. There are also other Feynman diagrams for this process.}
\label{fig:gamma}
\end{center}
\end{figure}

The cross section for annihilation of two neutralinos to two
photons has been calculated by a number of authors (e.g., 
Refs. \cite{rudaz,b89,gg89,gammas,piero,bern}).  
Annihilation can proceed, e.g., through a Feynman diagram
such as that shown in Fig. \ref{fig:gamma}, but there are also
many other diagrams that can contribute.  The value of the
annihilation cross section depends on a number of supersymmetric
couplings and sparticle masses, and it may span a range of
several orders of magnitude.  In some regions of the
plausible supersymmetric parameter space, it can indeed be as
large as $10^{-38}$ cm$^2$ \cite{piero}.

The diagrams for elastic scattering of photons from neutralinos
are related to these by crossing symmetry, and this led
B\oe hm et al. to speculate that the elastic-scattering cross
sections could also be as large as $10^{-38}$ cm$^2$.  This,
however, neglects a dramatic suppression of the
elastic-scattering cross section that arises from the spinor
structure of the amplitude.

The amplitude for either process can be written,
\begin{equation}
     {\cal A} = {e^2 \over 4 \pi^2} \varepsilon_{\mu\nu\rho\sigma}
     k^\mu k^\nu \epsilon^\rho \epsilon^\sigma \bar{\cal A},
\end{equation}
where $k^\mu$ and $k^\nu$ are the four-momenta of the photons
and $\epsilon^\rho$ and $\epsilon^\sigma$ are their polarization 
four-vectors and $\varepsilon_{\mu\nu\rho\sigma}$ is the completely
antisymmetric tensor.  The reduced amplitude $\bar{\cal A}$ depends
on the particle masses and couplings in the diagram loops.  Its
value for annihilation will differ from that for 
elastic scattering by factors of order unity, but not by many
orders of magnitude.  For the
annihilation process, the four-momenta
will be equal to $m_\chi$, while for elastic scattering,
they will be equal to the photon energy.  Thus, neglecting the
small difference in $\bar {\cal A}$, we see that the factors of
$k$ in the amplitude will lead to a reduction by a factor
$(E_\gamma/m_\chi)^4$ of the elastic-scattering cross section
relative to that for annihilation.  For $E_\gamma\sim0.1$ keV
and a neutralino mass $m_\chi\sim100$ GeV, this suppression is
$10^{-36}$.  More precisely, the annihilation cross
section is
\begin{equation}
     \sigma_{\chi\chi\rightarrow \gamma\gamma} = {\alpha^2
     m_\chi^2 \over 16 \pi^3} |\bar{\cal A}|^2,
\end{equation}
while the elastic-scattering cross section is
\begin{equation}
     \sigma_{\chi\gamma\rightarrow \chi\gamma} =
     \frac{\alpha^2}{3\pi^2 m_{\chi}^2} E_{\gamma}^4 |\bar{\cal
     A}|^2.
\end{equation}
We have calculated this cross section for a number of
supersymmetric models using a modified version of  
the code {\tt neutdriver} \cite{report}.  For a range of typical 
supersymmetric parameters  and a photon energy of 100 eV, we find
$10^{-79}\, {\rm cm}^2 < \sigma < 10^{-73}\, {\rm cm}^2 $. 
We thus conclude that elastic neutralino-photon scattering will
play no role in maintaining kinetic equilibrium of neutralinos,
nor of scattering photons from neutralinos.

\section{Neutralino-neutrino scattering}

\begin{figure*}[htbp]
\begin{center}
\begin{minipage}{0.28\textwidth}
\begin{center}
\epsfig{file=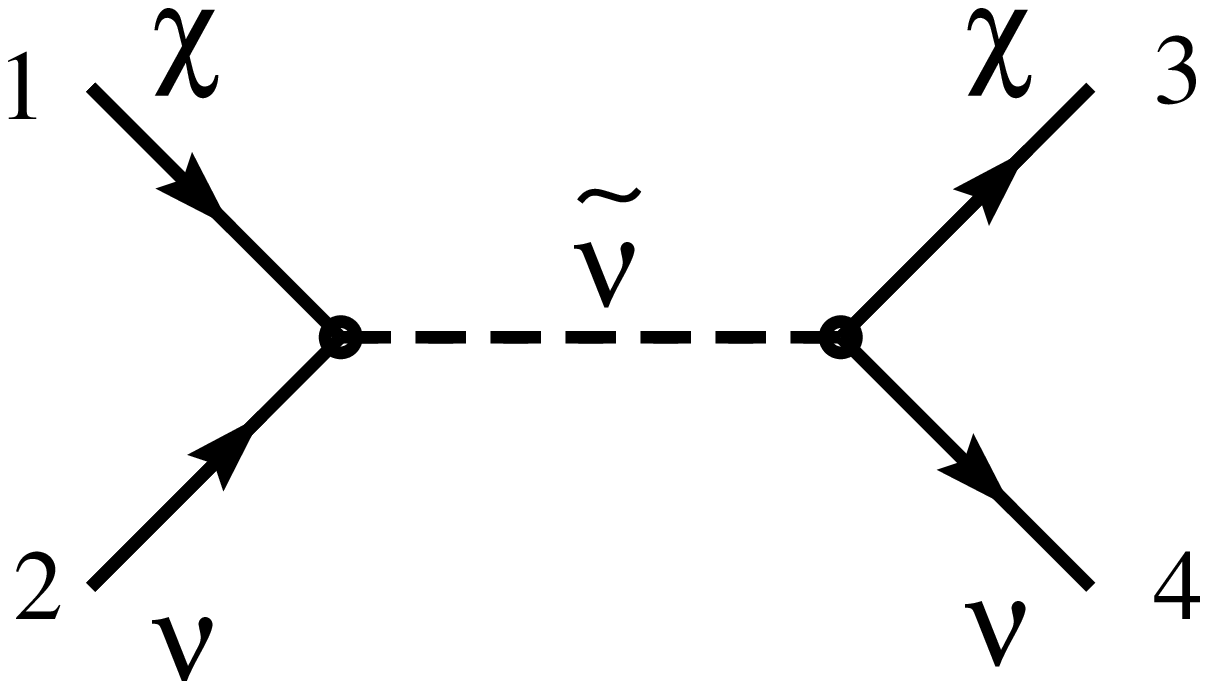,width=\textwidth}
\end{center}
\end{minipage}
\begin{minipage}{0.28\textwidth}
\begin{center}
\epsfig{file=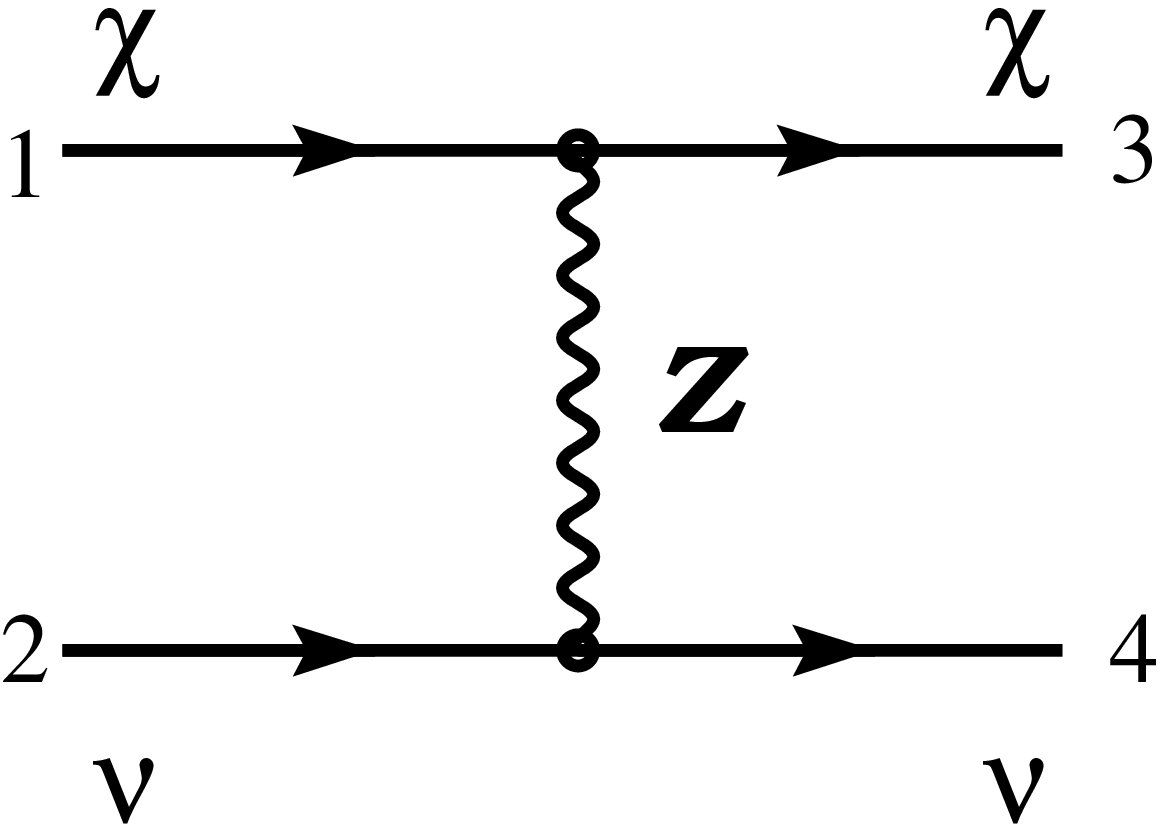,width=\textwidth}
\end{center}
\end{minipage}
\begin{minipage}{0.28\textwidth}
\begin{center}
\epsfig{file=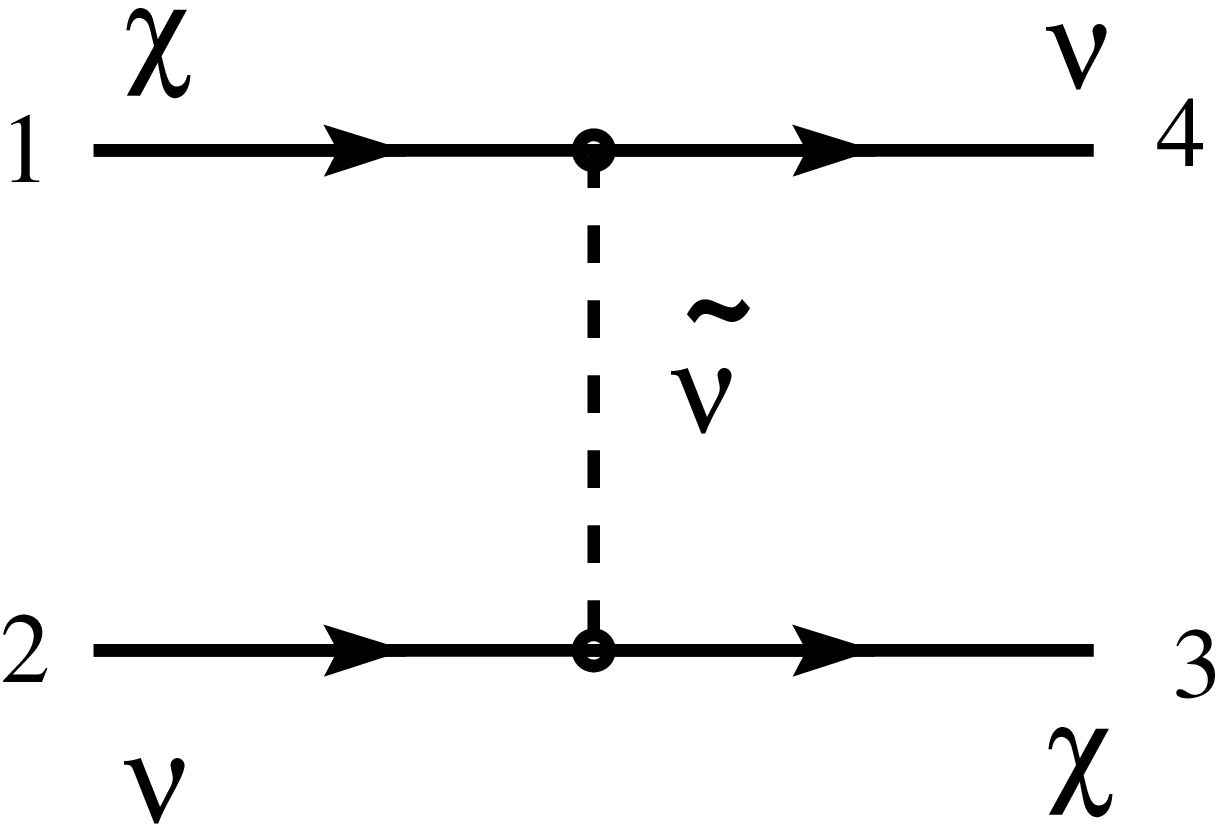,width=\textwidth}
\end{center}
\end{minipage}
\end{center}
\vspace{6pt}
\caption{Feynman diagrams for elastic neutralino-neutrino scattering.}
\label{fig:neut}
\end{figure*}

The neutrino-neutralino differential scattering cross section is
given by
\begin{equation}
     \left(\frac{d\sigma}{d\Omega}\right) =
     \frac{|{\mathcal M}|^2}{64\pi^2 m_{\chi}^2},
\end{equation}
where 
\begin{equation}
      {\mathcal M}={\mathcal M}_{s} +{\mathcal M}_{t}+{\mathcal M}_{u}
\end{equation}
is the scattering amplitude. The Feynman diagrams are shown
in Fig. \ref{fig:neut} (we have omitted Higgs-boson-exchange diagrams since 
the Yukawa coupling of the Higgs boson is zero in the limit of zero 
neutrino mass). These are given by\footnote{For Feynman rules of 
Majorana fermions, see, e.g., Ref.~\cite{HK}}
\begin{eqnarray}
     {\mathcal M}_{s}&=& i \frac{|X'_{\nu ij0}|^2}{s-m_{\tilde{\nu_j}}^2} 
     \overline{u}_4 P_R C \overline{u}_3^{T} u_1^T C^{-1} P_L u_2,\nonumber\\
     {\mathcal M}_{t}&=& i \frac{g^2}{2\cos^2\theta_w}\frac{1}{t-m_Z^2}
     \overline{u}_4 \gamma^{\mu} P_L u_2\nonumber \\
     & & \overline{u}_3 \gamma_{\mu} 
     \left(O_{00L}'' P_L + O_{00R}'' P_R\right) u_1,\nonumber\\
     {\mathcal M}_{u}&=& -i \frac{|X'_{\nu ij0}|^2}{u-m_{\tilde{\nu_j}}^2}
     \overline{u}_4 P_R u_1 \overline{u}_3 P_L u_2,
\end{eqnarray}
where $i,j$ are the flavor indices of the neutrino and sneutrino,
respectively; $P_L=(1-\gamma_5)/2$ and $P_R=(1+\gamma_5)/2$;
$X'_{\nu ij0}$ is the neutralino-sneutrino-neutrino coupling
(given in Ref. \cite{report}); and $O_{00R}''$ and $O_{00L}''$
are neutralino-$Z^0$ couplings (also given in
Ref. \cite{report}). The superscript $T$ denotes the transpose
of the matrix, and $C$ is the charge-conjugation matrix. Then, 
\begin{eqnarray}
     |{\mathcal M}|^2&=& |{\mathcal M}_{s}|^2 + |{\mathcal M}_{t}|^2 + 
     |{\mathcal M}_{u}|^2 \nonumber \\
     & &+ 2 \Re\left({\mathcal M}_{s}{\mathcal M}_{t}^{*}
     +{\mathcal M}_{s}{\mathcal M}_{u}^{*}
     +{\mathcal M}_{t}{\mathcal M}_{u}^{*}\right).
\end{eqnarray}
The various terms are given
\begin{eqnarray}
     |{\mathcal M}_{s}|^2 &=&  \frac{2|X'_{\nu ij0}|^4}
     {(s-m_{\tilde{\nu_j}}^2)^2}
     (p_1  p_2)(p_3  p_4),\\
     |{\mathcal M}_{t}|^2 &=& 
     \frac{2g^4}{\cos^4\theta_w (t-m_Z^2)^2} \nonumber \\
     & \times & [|O_{00L}''|^2 (p_1  p_2)( p_3  p_4) + 
     |O_{00R}''|^2 (p_1  p_4)( p_2  p_3) \nonumber \\
     & & - 
     O_{00L}''O_{00R}'' m_{\chi}^2 (p_2  p_4)],\\
     |{\mathcal M}_{u}|^2 &=&  \frac{2|X'_{\nu
     ij0}|^4}{(u-m_{\tilde{\nu_j}}^2)^2} 
     (p_1  p_4)(p_2  p_3),\\
     {\mathcal M}_{s} {\mathcal M}_{t}^*&=& -\frac{g^2}{\cos^2\theta_w}
     \frac{|X'_{\nu ij0}|^2}{s-m_{\tilde{\nu_j}}^2}
     \frac{1}{t-m_{\tilde{\nu_j}}^2} \nonumber \\
     & & \times\left[2 O_{00L}''^{*} 
     (p_1  p_2)( p_3  p_4)- O_{00R}''^{*}
     m_{\chi}^2 (p_2  p_4 ) \right],\\
     {\mathcal M}_{s} {\mathcal M}_{u}^*&=& - 
     \frac{|X'_{\nu ij0}|^4}{(s-m_{\tilde{\nu_j}}^2)(u-m_{\tilde{\nu_j}}^2)}
     m_{\chi}^2 (p_2 p_4),\\
     {\mathcal M}_{t} {\mathcal M}_{u}^*&=& 
     - \frac{g^2}{\cos^2\theta_w}\frac{|X'_{\nu ij0}|^2}
     {(t-m_Z^2)(u-m_{\tilde{\nu_j}}^2)} \nonumber \\
     & &  \times \left[2O_{00R}'' (p_1  p_4)
     (p_2  p_3) -  O_{00L}'' m_{\chi}^2 (p_2  p_4) \right].
\end{eqnarray}

In the limit $m_{\chi} \gg E_{\nu}$, we have
\begin{equation}
     s \simeq m_{\chi}^2, \qquad 
     t \simeq -2 (p_2  p_4), \qquad  
     u \simeq m_{\chi}^2,
\end{equation}
and
\begin{eqnarray}
     && (p_1  p_2) \simeq (p_2  p_3) \simeq m_{\chi}
     E_\nu,\nonumber\\
     && (p_1  p_4) \simeq (p_3  p_4) \simeq m_{\chi} E_\nu,\\
     && (p_2  p_4) \simeq E_\nu^2 (1-\cos\theta).\nonumber
\end{eqnarray}
We thus see that for $m_\chi \gg E_\nu$, the square of the
matrix element, and thus the cross section, are proportional to
$(E_\nu/m_\chi)^2$, and so they are enormously suppressed as the 
neutrinos cool.  The cross section is thus, schematically,
\begin{equation}
     \sigma_{\rm el} \sim ({\rm couplings})^2
     (E_\nu^2/m_\chi^4),
\label{eq:elastic}
\end{equation}
and should be at most (for couplings of order unity)
$10^{-46}$ cm$^2$.  A
survey of the supersymmetric parameter space shows that when
realistic couplings are included, the typical cross section is more like
$10^{-53}$ cm$^2$.  The cross section can be enhanced
if the sneutrino mass is nearly degenerate with the neutralino
mass in which case there can be an $s$- or $u$-channel
resonance.  However, the sneutrino would have to be
extraordinarily close in mass to the neutralino, and there is
no reason why this should be so in supersymmetric models.

\section{Kinetic Decoupling More Generally}

More generically, the annihilation cross section should be,
\begin{equation}
     \sigma_{\rm ann} \sim ({\rm couplings})^2 /m_\chi^2,
\end{equation}
and this must be $\sim10^{-38}$ cm$^2$ if the neutralino is to be the dark 
matter.  This refers to the cross section for annihilation to
light quarks, leptons, and neutrinos.   The elastic-scattering
cross section will be related to this by crossing symmetry, but
with the correct kinematic factor, we obtain an elastic cross
section like Eq. (\ref{eq:elastic}) (but with $E_\nu$ replaced
more generally by the light-fermion energy).
The rate at which a given neutralino will scatter from
these light particles will be
$\Gamma_{\rm el} \sim n_f \sigma_{\rm el}$, where $n_f\sim T^3$ is
the light-particle density, $T$ is the temperature, and at this
temperature, the elastic-scattering cross section will be
$\sigma_{\rm el} \sim 10^{-37}\, {\rm cm}^2\,(T/m_\chi)^2$.
Equating this elastic-scattering rate to the expansion rate,
$H\sim T^2/m_{\rm Pl}$, we find that neutralinos should undergo
kinetic decoupling at
\begin{equation}
     T_{\rm kin~dec} \sim {\rm MeV}\, (m_\chi/{\rm GeV})^{2/3}.
\end{equation}
Thus, after the neutralino freezes out at a temperature $T\sim
m_\chi/20$, it will remain in kinetic equilibrium with the
plasma and then undergo kinetic decoupling shortly thereafter.
For neutralino masses $10 \, {\rm GeV} \lesssim m_\chi \lesssim
1000\, {\rm GeV}$, this takes place well before neutrino
decoupling and electron-positron annihilation.

At temperatures $T\lesssim100$ MeV, the baryons will be in the
form of neutrons and protons, and shortly thereafter, 25\% of
the mass will be bound up in helium nuclei.  The
neutralino-nucleon cross section required to maintain neutralino
kinetic equilibrium at a temperature $T\sim$MeV is roughly
$10^{-35}$ cm$^2$.  This cross section is extremely high for
supersymmetric models (typical cross sections are
$\lesssim10^{-43}$ cm$^2$), and is in fact ruled out by null
searches for energetic neutrinos from WIMP annihilation in the
Sun and Earth\cite{ukv}.

\section{Conclusions}

We have shown that the cross sections for elastic scattering of
neutrinos and photons from neutralinos are far too small to keep 
neutralinos in kinetic equilibrium until $T\sim10-100$ eV, when
the Hubble length encloses dwarf-galaxy masses.  Moreover, the
cross sections required for collisional damping may be ten
orders of magnitude bigger, in which case neutrino and photon
elastic scattering are that much further from being able to
provide collisional damping on these scales.  Therefore, elastic 
scattering of neutralinos from neutrinos or photons cannot erase 
structure on sub-galactic scales.

We have moreover argued that neutralinos should generically
decouple kinetically from the primordial plasma shortly after
they cease annihilating, above temperatures $T\sim$MeV (at which 
time a Hubble volume encloses no more than a solar mass of
baryons).  Neutralinos will thereafter act as collisionless cold
dark matter and thus be incapable of altering the power spectrum
on scales relevant for galaxy and large-scale-structure formation.

\acknowledgments 
X.C. acknowledges the hospitality of the TAPIR group at Caltech where part of
this work was completed. X.C. was supported in part by the 
U.S. Department of Energy under grant No. DE-FG02-91ER40690. M.K. was
supported by the DoE under DE-FG03-92-ER40701, the NSF under
AST-0096023, and NASA under NAG5-8506.  X.Z. was supported in
part by the National Natural Science Foundation of China and the
Ministry of Science and Technology of China under grant
No. NKBRSF G19990754.

{\it Note added to proof:} Shortly after the publication of our
preprint, another paper \cite{hs} which also discussed kinetic
decoupling of cold dark matter came out.

\end{document}